\documentclass[aps,prd,nofootinbib]{revtex4}
\pdfoutput=1

\usepackage{amsmath}
\usepackage{amssymb}
\usepackage{anysize}
\usepackage{bold-extra}
\usepackage{color}
\usepackage{enumerate}
\usepackage{fancyhdr}
\usepackage{graphicx}
\usepackage[linktocpage,colorlinks,urlcolor=blue,citecolor=blue]{hyperref}
\usepackage{lscape}
\usepackage{mathrsfs}
\usepackage{multirow}
\usepackage[final]{pdfpages}
\usepackage{rotating}
\usepackage{subfig}
\usepackage{textcomp}
\usepackage{url}
\usepackage{verbatim}
\usepackage{wrapfig}
\usepackage{xcolor}
\usepackage{amsfonts}
\usepackage{comment}
\usepackage{epigraph}
\usepackage[utf8]{inputenc}
\usepackage{slashed}
\usepackage{bbm}
\usepackage{cancel}
\usepackage{epsf, array, color}
\def\nn{\nonumber}

\begin{document}

\title{Model-Independent Determination of $B_c^+ \to \eta_c\, \ell^+\, \nu$ Form Factors }

\author{Christopher W. Murphy}
\email{chrismurphybnl@gmail.com}
\affiliation{Department of Physics, Brookhaven National Laboratory, Upton, N.Y., 11973, U.S.A.}

\author{Amarjit Soni}
\email{adlersoni@gmail.com}
\affiliation{Department of Physics, Brookhaven National Laboratory, Upton, N.Y., 11973, U.S.A.}

\begin{abstract}
We derive model-independent bounds on the form factors for the decay $B_c^+ \to \eta_c\, \ell^+\, \nu$ including full mass effects, \textit{i.e.} $\ell = e,\, \mu, \text{ and } \tau$.
The bounds are obtained by using the BGL parameterization for the form factors, and fitting to the preliminary lattice data of the HPQCD Collaboration.
Our main result after bounding the form factors is the Standard Model (SM) prediction for the ratio of branching fractions $R(\eta_c) = \mathcal{B}(B_c^+ \to \eta_c\, \tau^+\, \nu_{\tau}) / \mathcal{B}(B_c^+ \to \eta_c\, \mu^+\, \nu_{\mu})$.
We find $R(\eta_c)|_{\text{SM}} = 0.31^{+0.04}_{-0.02}$, and argue that a measurement of $R(\eta_c)$ is within the reach of LHCb during the high luminosity run of the LHC.
In addition, using the heavy-quark spin symmetry of the $B_c$ meson we relate our results for $B_c^+ \to \eta_c\, \ell^+\, \nu$ to those for $B_c^+ \to J/\psi\, \ell^+\, \nu$ yielding the estimate $R(J/\psi)|_{\text{SM}} = 0.26 \pm 0.02$ in good agreement with other determinations.
\end{abstract}

\maketitle

\section{Introduction}
\label{sec:in}
Measurements by the BaBar~\cite{Lees:2012xj, Lees:2013uzd}, Belle~\cite{Huschle:2015rga, Hirose:2016wfn, Hirose:2017dxl}, and LHCb~\cite{Aaij:2015yra, Aaij:2017uff, Aaij:2017deq} experiments are challenging the assumption of lepton universality in the decays of $B$-mesons.
In particular, the ratios of branching fractions
\begin{equation}
R(D^{(*)}) \equiv \frac{\mathcal{B}(B \to D^{(*)} \tau \nu_{\tau})}{\mathcal{B}(B \to D^{(*)} l \nu_{l})} 
\end{equation}
are measured to be larger than predicted in the Standard Model (SM).\footnote{In this work $l = e,\, \mu$; $\ell = e,\, \mu,\, \tau$; and charge-conjugation is implied throughout.}
A global analysis~\cite{HFLAV:website} quantifies the discrepancy as a $\sim3.8\sigma$ deviation from the SM predictions~\cite{Fajfer:2012vx, Lattice:2015rga, Na:2015kha, Bigi:2016mdz, Bernlochner:2017jka, Bigi:2017jbd, Jaiswal:2017rve}.
These measurements are complemented by tests of lepton universality by LHCb in the decays $B \to K^{(*)} l^+ l^-$, which show deficits with respect to the SM predictions~\cite{Aaij:2014ora, Aaij:2017vbb}.
If confirmed, either of the charged- or neutral-current measurements would be unambiguous signals of the breakdown of the SM.
See Refs.~\cite{Barbieri:2016las, Amhis:2016xyh, Ciezarek:2017yzh, Altmannshofer:2017poe, Azatov:2018knx} for some overviews of the situation.

If the enhancement of $R(D^{(*)})$ is due to physics beyond the SM it should be evident in other decays of $B$-hadrons as well, such as $\Lambda_b \to \Lambda_c\, \ell\, \nu$ and $B_c \to (c\bar{c})\, \ell\, \nu$, as the underlying weak matrix elements for $b \to c \ell \nu$ are independent of the hadronic physics.
In fact, LHCb has also measured~\cite{Aaij:2017tyk} a larger than expected value for the ratio
\begin{equation}
R(J/\psi) \equiv \frac{\mathcal{B}(B_c^+ \to J/\psi\, \tau^+\, \nu_{\tau})}{\mathcal{B}(B_c^+ \to J/\psi\, \mu^+\, \nu_{\mu})} .
\end{equation}
However until very recently there was no model-independent prediction for $R(J/\psi)$.
As this work was being finalized Ref.~\cite{Cohen:2018dgz} appeared, which used the Boyd, Grinstein, Lebed (BGL) parameterization~\cite{Boyd:1994tt, Boyd:1995sq, Boyd:1997kz} to bound the form factors for $B_c^+ \to J/\psi\, \ell^+\, \nu$ in a model-independent fashion.
In light of this recent development we will concentrate on the complementary decay $B_c^+ \to \eta_c\, \ell^+\, \nu$ in this work.

Currently there is no measurement of any decay $B_c^+ \to \eta_c\, X^+$, let alone a measurement of the analogous $R$ ratio involving an $\eta_c$ meson
\begin{equation}
R(\eta_c) \equiv  \frac{\mathcal{B}(B_c^+ \to \eta_c\, \tau^+\, \nu_{\tau})}{\mathcal{B}(B_c^+ \to \eta_c\, \mu^+\, \nu_{\mu})} .
\end{equation}
As no measurement currently exists, it is natural to wonder when a measurement of a $B_c^+ \to \eta_c\, X^+$ process might be made.
In what follows we argue that a measurement of $R(\eta_c)$ is within the reach of LHCb during the planned high-luminosity run of the LHC (HL-LHC).
LHCb has observed $\eta_c$ in the decay $B_s^0 \to \eta_c\, \phi$, and has evidence for the decay $B_s^0 \to \eta_c\, \pi^+\, \pi^-$~\cite{Aaij:2017hfc}.
The branching fractions for these $\eta_c$ modes are comparable to the corresponding decays involving a $J/\psi$~\cite{Tanabashi:2018oca}.
Furthermore, the measured branching fraction for $B_c^+ \to J/\psi\, \ell^+\, \nu_{\ell}$ is similar to the other measured charmonium mode $B_c \to \chi_{c0}\, \pi^+$~\cite{Aaij:2016xas}.
These similarities suggest the main difference in the rates for $B_c^+ \to \eta_c\, \ell^+\, \nu_{\ell}$ and $B_c^+ \to J/\psi\, \ell^+\, \nu_{\ell}$ is in the branching fractions for what the charmonia decay into.
Ref.~\cite{Aaij:2017tyk} used the decay mode $J/\psi \to \mu^+ \mu^-$ to measure $R(J/\psi)$, which has a well known branching fraction of slightly less than 6\%.
On the other hand, the main decay modes LHCb used to observe the $\eta_c$ in $B_s^0$ decays were $p \bar{p}$, $\pi^+ \pi^- \pi^+ \pi^-$, and $K^+ K^- K^+ K^-$, which combine to give a total branching fraction of approximately 1\%~\cite{Tanabashi:2018oca}.
All else being equal this implies if these same modes were used by LHCb to measure $R(\eta_c)$ then roughly 100~fb$^{-1}$ of integrated luminosity would be needed to match the statistics of $R(J/\psi)$ analysis, which used the full Run-1 dataset of 3~fb$^{-1}$.
This is within the 300~fb$^{-1}$ target for LHCb at HL-LHC.
Furthermore, it is possible that adding a cleaner, but rarer mode, such as $\eta_c \to \gamma \gamma$, could improve the efficiency and lower the amount of the data required.

Confident that a measurement of $R(\eta_c)$ will eventually be made, we derive model-independent bounds on the form factors for $B_c^+ \to \eta_c\, \ell^+\, \nu$ using the BGL parameterization.
This starts in Sec.~\ref{sec:ff} by reviewing the BGL formalism, and applying it to the case at hand, $B_c^+ \to \eta_c\, \ell^+\, \nu$.
Sec.~\ref{sec:lqcd} describes the lattice data from the HPQCD collaboration that is used to fit the form factors.
Our results for the form factors are given in Sec.~\ref{sec:res}.
We find $R(\eta_c)|_{\text{SM}} = 0.31^{+0.04}_{-0.02}$.
Then in Sec.~\ref{sec:hqss} we use heavy-quark spin symmetry (HQSS) to relate the form factors for $B_c^+ \to \eta_c\, \ell^+\, \nu$ to those for $B_c^+ \to J/\psi\, \ell^+\, \nu$, yielding the estimate $R(J/\psi)|_{\text{SM}} = 0.26 \pm 0.02$.
Concluding remarks follow in Sec.~\ref{sec:con}.

\section{Form Factors}
\label{sec:ff}
The hadronic matrix element describing the decay $B_c^+ \to \eta_c\, \ell^+\, \nu$ contains two factors.
It can be written analogously to other form factors for pseudoscalar-to-pseudoscalar transitions
\begin{equation}
\label{eq:matel}
\langle \eta_c(p^{\prime}) | V^{\mu} | B_c(p) \rangle = f_+(q^2) (p + p^{\prime})^{\mu} + \left(f_0(q^2) - f_+(q^2)\right) \frac{M_{B_c}^2 - M_{\eta_c}^2}{q^2} (p - p^{\prime})^{\mu},
\end{equation}
where $V^{\mu} = \bar{c} \gamma^{\mu} b$ and $q^2 = (p - p^{\prime})^2$.
With this definition we must have $f_0(0) = f_+(0)$ in order for the matrix element in Eq.~\eqref{eq:matel} to be finite at $q^2 = 0$.\footnote{In contrast zero-recoil is defined by $q^2 = q^2_{\text{max}}$, where both the initial and final mesons are at rest.}
The differential decay rate is then given by
\begin{equation}
\label{eq:gamma}
\frac{d\Gamma}{dq^2}(B_c^+ \to \eta_c\, \ell^+\, \nu) = \frac{\eta_{ew}^2 G_F^2 |V_{cb}|^2 M_{B_c} \sqrt{\lambda}}{192 \pi^3} \left(1 - \frac{m_{\ell}^2}{q^2}\right)^2 \left[c_+ f_+(q^2)^2 + c_0 f_0(q^2)^2 \right]
\end{equation}
with $\lambda \equiv \lambda(q^2, M_{B_c}^2, M_{\eta_c}^2)$, where $\lambda(a, b, c) = a^2 + b^2 + c^2 - 2 (a b + b c + c a)$, and
\begin{equation}
c_+ = \frac{\lambda}{M_{B_c}^4} \left(1 + \frac{m_{\ell}^2}{2 q^2}\right), \quad c_0 = (1 - r^2)^2 \frac{3 m_{\ell}^2}{2 q^2}, \quad r = \frac{M_{\eta_c}}{M_{B_c}} .
\end{equation}

We use the BGL parameterization~\cite{Boyd:1994tt, Boyd:1995sq, Boyd:1997kz} in determining the form factors $f_+$ and $f_0$.
The BGL formalism follows from analyticity, crossing symmetry, and dispersion relations.
This leads to unitarity bounds on the otherwise free parameters that characterize the form factors.
These properties follow from QCD, and are independent of any model.\footnote{The numerical values for the masses of the $B_c$ scalar and vector mesons are not yet measured experimentally. However they are well determined on the lattice~\cite{Dowdall:2012ab}.}
For pedagogical introductions see Refs.~\cite{Boyd:1995tg, Boyd:1997qw}.
The final result is that the form factors are written as a Taylor series
\begin{equation}
\label{eq:ff}
f_i(z) = \frac{1}{P_i(z) \phi_i(z)} \sum_{i; n = 0}^{\infty} a_{i, n} z^n ,
\end{equation}
for $i = \{+, 0\}$, and the parameters $a_{i,n}$ are bounded by unitarity to satisfy
\begin{equation}
\label{eq:uni}
\sum_{n = 0}^{\infty} a_{+, n}^2 < 1 , \quad \sum_{m = 0}^{\infty} a_{0, m}^2 < 1.
\end{equation}

Having presented the final result, we now describe the ingredients of the BGL parameterization.
We begin by defining $t = q^2$ and $t_{\pm} = (M_{B_c} \pm M_{\eta_c})^2$.
A conformal transformation is made, $t \to z$, which maps the $t$ plane to the unit disk.
There is freedom to choose the point that is mapped to $z = -1$.
We take this to be $t_* = (M_{B_c} + M_{\eta})^2 \to -1$ such that there are no isospin preserving branch cuts within the unit disk.\footnote{Note that our branch point involves the mass of the $\eta$ and not the $\eta_c$. For an axial-vector current the first branch cut occurs for the three-body process $B_c + 2 \pi$. Although branch cuts arising from $B_c + \text{ light hadrons}$, and cuts with $\geq 3$ particles, are dynamically suppressed with respect to those where each hadron contains a heavy quark, \textit{e.g.} $B^{(*)} + D$~\cite{Boyd:1995tg}, the freedom exists to choose the branch point.}
There is further freedom in choosing the point that is mapped to the origin.
Here we make the most common choice in the literature, $t_0 = t_- \to 0$, and do not  optimize this value.
With these choices the conformal parameter is given by
\begin{equation}
z(t; t_0) = \frac{t_0 - t}{(\sqrt{t_* - t} + \sqrt{t_* - t_0})^2} ,
\end{equation}
The minimal value of $z$ is $z_{\text{min}} = z(t_-; t_0) = 0$, whereas the maximal value is given by $z_{\text{max}} = z(m_{\ell}^2; t_0)$ leading to $z_{\text{max}, \tau} \approx 0.049$ and $z_{\text{max}, \mu} \approx z_{\text{max}, e} \approx 0.066$.
See Table~\ref{tab:mass} for the numeric values employed in this work. 
Due to the small range of $z$, the series in Eq.~\eqref{eq:ff} can be safely truncated after only a few terms.

The pre-sum factors appearing in Eq.~\eqref{eq:ff}, $P$ and $\phi$, are known as Blaschke factors and outer functions, respectively.
A Blaschke factor removes all the poles due single-particle states appearing below the pair-production threshold $t_*$
\begin{equation}
\label{eq:blas}
P_i(z) = \prod_{p_i} \frac{z \cdot z_{p_i}}{1 - z \cdot z_{p_i}},
\end{equation}
where $z_{p_i} = z(M_{p_i}^2; t_0)$, again for $i = \{+, 0\}$.
An advantage of our choice for $t_*$ is each form factor only has one pole within the unit disk.
The masses of the relevant $B_c$ states, $0^+$ and $1^-$, are given in Table~\ref{tab:mass}.
The numerical values for these masses were taken from Ref.~\cite{Dowdall:2012ab}.
\begin{table}
\centering
 \begin{tabular}{ c | c | c }
Particle & Mass [GeV] & Reference \\ \hline \hline
$B_c(0^-)$ & 6.2749 & \cite{Tanabashi:2018oca} \\ \hline
$\eta_c$ & 2.9839 & \cite{Tanabashi:2018oca} \\ \hline
$\eta$ & 0.547862 & \cite{Tanabashi:2018oca} \\ \hline
$J/\psi$ & 3.09690 & \cite{Tanabashi:2018oca} \\ \hline
$\tau$ & 1.77686 & \cite{Tanabashi:2018oca} \\ \hline 
$\mu$ & 0.105658 & \cite{Mohr:2015ccw} \\ \hline \hline
$B_c(1^-)$ & 6.329(8) & \cite{Dowdall:2012ab} \\ \hline
$B_c(0^+)$ & 6.704(17) &  \cite{Dowdall:2012ab} \\ \hline \hline
$b$ & 4.163(16) & \cite{Chetyrkin:2009fv} \\ \hline
$c$ & 0.986(13) & \cite{Chetyrkin:2009fv} \\ 
 \end{tabular}
  \caption{Numeric values for the masses employed in this work.
  The first six masses are experimentally measured.
  The masses of scalar and vector $B_c$ mesons were determined on the lattice, and only enter the form factors through the Blaschke factors appearing in Eq.~\eqref{eq:blas}.
  The bottom- and charm-quark masses are given in the $\overline{\text{MS}}$ scheme at $\mu = m_b$ and $\mu = 3$~GeV, respectively.
  They only explicitly enter our calculations through the heavy-quark spin symmetry relations given in Sec.~\ref{sec:hqss}, but are also implicitly used in the evaluations of the $\chi$ functions in Eq.~\eqref{eq:chi0}.}
  \label{tab:mass}
\end{table}
The outer functions depend on kinematics and parameters $\chi$ that appear in the dispersion relations
\begin{align}
\label{eq:disp}
\chi^L(q^2) &\equiv \frac{\partial \Pi^L}{\partial q^2}= \frac{1}{\pi} \int_0^{\infty} \! dt\, \frac{\text{Im } \Pi^L(t)}{(t - q^2)^2} , \\
\chi^T(q^2) &\equiv \frac{1}{2} \frac{\partial^2 \Pi^T}{\partial (q^2)^2} = \frac{1}{\pi} \int_0^{\infty} \! dt\, \frac{\text{Im } \Pi^T(t)}{(t - q^2)^3} , \nn
\end{align}
where
\begin{equation}
\frac{1}{q^2} (q^{\mu} q^{\nu} - q^2 g^{\mu\nu}) \Pi^T(q^2) + \frac{q^{\mu} q^{\nu}}{q^2} \Pi^L(q^2) = i \int \! d^4x \, e^{i q \cdot x} \langle 0 | \text{T } V^{\mu}(x) V^{\dagger \nu}(0) | 0 \rangle .
\end{equation}
Note that the integrands appearing on the right-hand sides of Eqs.~\eqref{eq:disp} must be analytic below the pair-production threshold, which in this case is $t_* = (M_{B_c} + M_{\eta})^2$, if the dispersion relations are to be used to constrain the form factors in the semileptonic region.
This is why it is necessary to include the Blaschke factors and judiciously choose $t_*$.
The $\chi$ can be perturbatively computed in QCD for $q^2$ near 0.
The state-of-the-art three-loop evaluations are~\cite{Bigi:2016mdz}
\begin{equation} 
\label{eq:chi0}
\chi^T(0) = 6.486(48) \cdot 10^{-4} \text{ GeV}^{-2}, \quad \chi^L(0) = 6.204(81) \cdot 10^{-3} .
\end{equation}
With $\chi$ in hand, the outer functions are given by~\cite{Boyd:1997kz}
\begin{align}
\phi_+(z) &= \frac{8}{M_{B_c}} \sqrt{\frac{8}{3 \pi \chi^T(0)}} \frac{r^2 \sqrt{1-z} (1 + z)^2}{((1 + r) (1 - z) + 2 \sqrt{r} (1 + z))^5} , \\
\phi_0(z) &= \sqrt{\frac{8}{\pi \chi^L(0)}} \frac{r (1 - r)^2 \sqrt{1 - z} (1 - z^2)}{((1 + r) (1 - z) + 2 \sqrt{r} (1 + z))^4} . \nn
\end{align}
This completes the list of ingredients needed to parameterize the form factors.

\section{Lattice QCD Calculations}
\label{sec:lqcd}
The HPQCD collaboration made preliminary predictions for both form factors $f_+$ and $f_0$ at four and five values of $q^2$, respectively, using Lattice QCD~\cite{Colquhoun:2016osw}.
In addition, in the same work preliminary predictions were made for two of four form factors describing the decay $B_c^+ \to J/\psi\, \ell^+\, \nu$.
The computations were done using a highly improved staggered quark action with $n_f = 2 + 1 + 1$ flavors and multiple lattice spacings.
HPQCD only reports statistical errors in Ref.~\cite{Colquhoun:2016osw}.
Following Ref.~\cite{Cohen:2018dgz} we assign a 20\% systematic uncertainty to these predictions to take into account discretization errors, finite-volume corrections, and quark-mass dependencies. 
This uncertainty is 20\% with respect to the central value of the prediction, and is added in quadrature with the corresponding statistical uncertainty.

\section{Results for $B_c^+ \to \eta_c\, \ell^+\, \nu$}
\label{sec:res}
In this section we give our results for the form factors $f_+$ and $f_0$ by fitting to the preliminary lattice data using the BGL parameterization.
We then proceed to calculate the SM value for the ratio $R(\eta_c)$.
We find the best-fit values for the parameters, $a_{i,n}$, by minimizing a chi-squared function for each form factor
\begin{equation}
\chi_+^2 = \sum_{j = 1}^{4} \frac{[f_+^{lat}(q_j^2) - f_+(q_j^2)]^2}{[\sigma_+^{lat}(q_j^2)]^2}, \quad \chi_0^2 = \sum_{j = 1}^{5} \frac{[f_0^{lat}(q_j^2) - f_0(q_j^2)]^2}{[\sigma_0^{lat}(q_j^2)]^2} .
\end{equation}
The superscript \textit{lat} indicates a lattice measurement of a form factor, $f$, and the corresponding uncertainty, $\sigma$, at a given value of $q^2$.
The $f$ without a superscript is the form factor to be fit from Eq.~\eqref{eq:ff}, which is subject to the unitarity bound Eq.~\eqref{eq:uni}.
The uncertainties on the form factors are found by propagating forward the correlated uncertainties on the $a$ parameters
\begin{equation}
\label{eq:deltaf}
\delta f_i^2(q^2) = \left(\nabla_{a_i} f_i\right) \cdot \left(\frac{1}{2} \frac{\partial^2 \chi^2_i}{\partial a_{i,n} \partial a_{i,m}}\right)^{-1} \cdot \left(\nabla_{a_i} f_i\right) .
\end{equation}
Note that because the form factors only have a linear dependence on the parameters, $a_i$ the uncertainties on the form factor are independent of the parameters themselves.

\begin{figure}
  \centering
\includegraphics[width=0.7\textwidth]{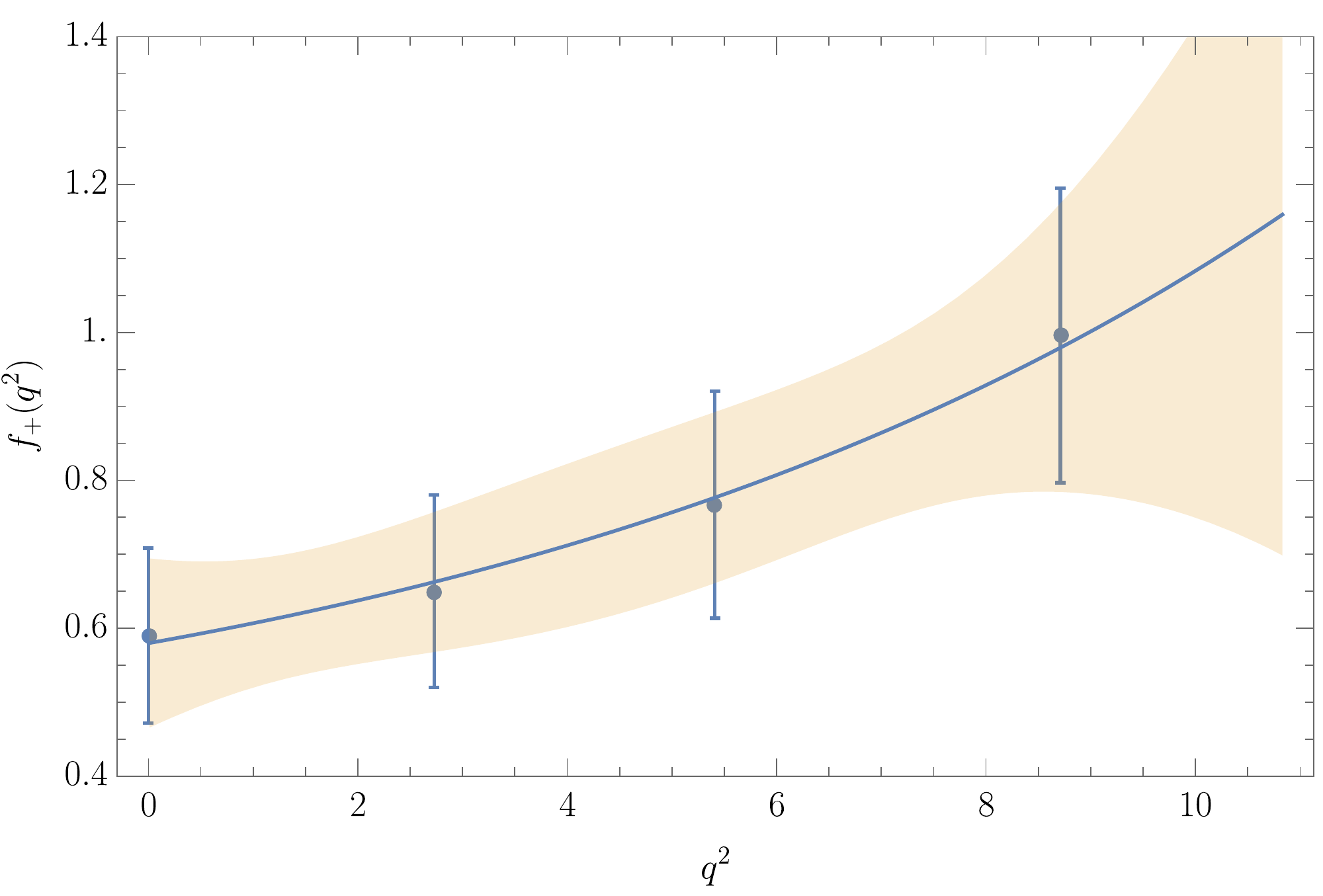}
 \caption{Result of our three parameter fit of the BGL-parameterized form factor, $f_+$, to the lattice data of Ref.~\cite{Colquhoun:2016osw}, which are the blue points with errors bars. 
 The blue curve is the best-fit value, and the shaded region is the $1\sigma$ allow range.}
  \label{fig:fp}
\end{figure}

\begin{figure}
  \centering
\includegraphics[width=0.7\textwidth]{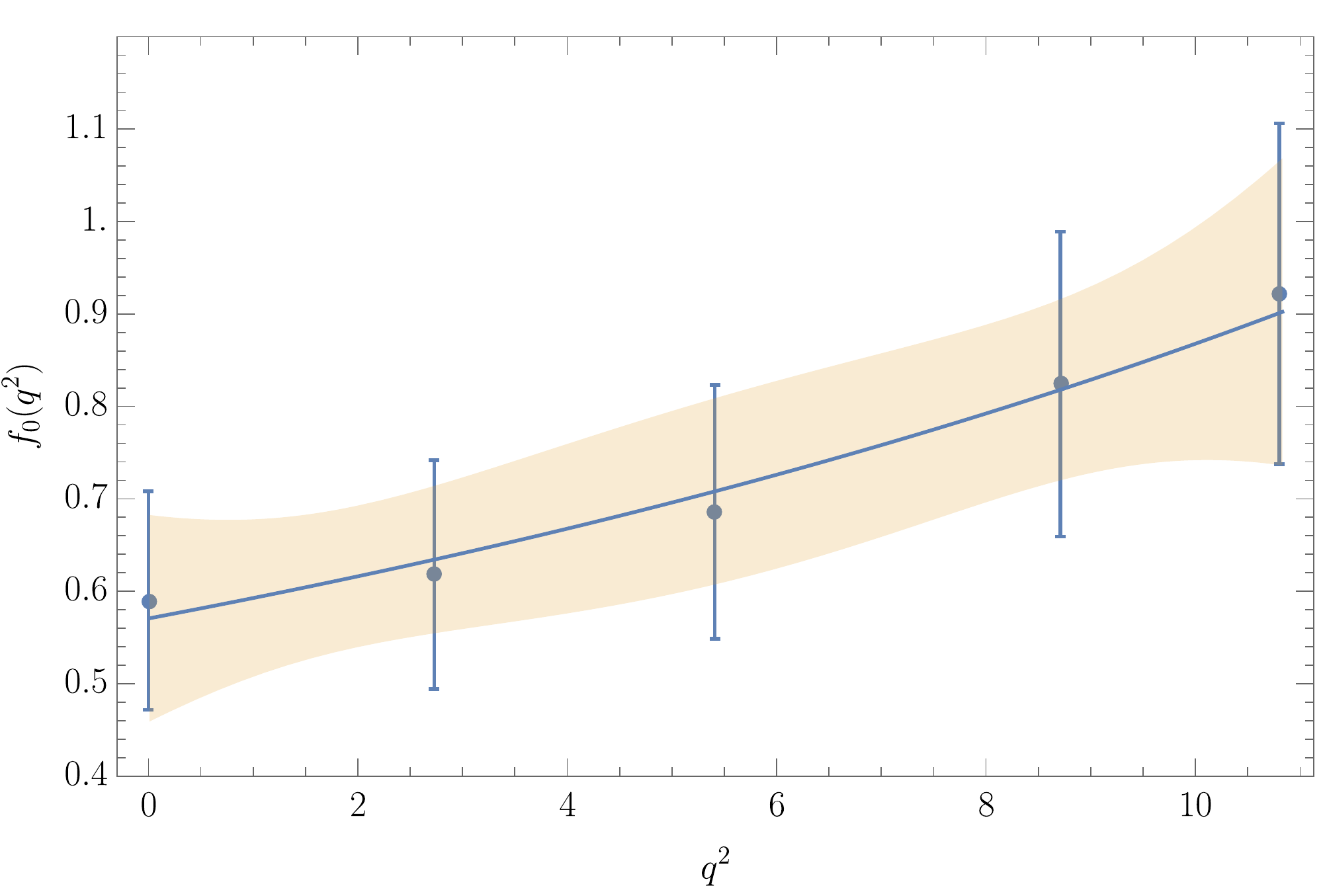}
 \caption{The same as Fig.~\ref{fig:fp}, but for the form factor $f_0$.}
  \label{fig:f0}
\end{figure}

We retain only the first three terms in Eq.~\eqref{eq:ff} for each form factor leading to maximum truncation errors of 1.0\% and 0.4\% for $f_+$ and $f_0$, respectively, both of which are well below the size of the lattice uncertainties.\footnote{The truncation error of an $n$ parameter fit is given by $\left(\Delta f_i\right)_{\text{trunc.}} = z_{\text{max}, \mu}^n / \left(P_i(z_{\text{max}, \mu}) \phi_i(z_{\text{max}, \mu})\right)$.}
In Fig.~\ref{fig:fp} and Fig.~\ref{fig:f0} we present the results of our three parameter fits for the form factors $f_+$ and $f_0$, respectively.
The lattice data from Ref.~\cite{Colquhoun:2016osw} are the blue points with error bars.
The blue curves are the best-fit values, and the shaded regions are the $1\sigma$ allow ranges.
The minimal value of $q^2$ is set by the mass of the muon in Figs.~\ref{fig:fp} and~\ref{fig:f0}.
Comparing the results of our fits for $f_i(B_c \to \eta_c)$ against those for $f_i(B \to D)$~\cite{Bigi:2016mdz} and the preliminary results for $f_i(B_s \to D_s)$~\cite{Witzel:2018} shows that the dependence of the form factors on the flavor of the spectator quark is mild.

The results of our fits can be found in Table~\ref{tab:res}.
\begin{table}
\centering
 \begin{tabular}{ c | c | c | c | c | c | c | c }
 & \multicolumn{3}{c |}{$f_+$} & \multicolumn{4}{c}{$f_0$} \\ \hline \hline
$n_{\text{params.}}$ & 1 & 2 & 3 & 1 & 2 & 3 & 4 \\ \hline
$\chi^2$ & 1.60 & 0.10 & 0.028 & 2.97 & 0.14 & 0.084 & 0.084 \\ \hline\hline
$a_{i, 0}$ & 0.018 & 0.024 & 0.026 & 0.051 & 0.065 & 0.066 & 0.066 \\ \hline
$a_{i, 1}$ &  & -0.13 & -0.21 &  & -0.35 & -0.41 & -0.41 \\ \hline
$a_{i, 2}$ &  &  & 0.98 &  &  & 0.91 & 0.90 \\ \hline
$a_{i, 3}$ &  &  &  &  &  &  & 0.10 \\ 
 \end{tabular}
  \caption{The results of our fits. The upper rows give the number of parameters and the $\chi^2$ for a given fit to $f_+$ or $f_0$. The bottom rows show the central values of the coefficients $a_i$ for a given fit.}
  \label{tab:res}
\end{table}
We checked that the goodness-of-fit is better for these three parameter fits than for one or two parameter fits even when the unitarity bounds are imposed.
For $f_0$ a four parameter fit is also possible, but it has a worse goodness-of-fit than the three parameter fit as well.
The reason is the unitarity bound becomes saturated once the third parameter is added (this is true in both the $f_+$ and $f_0$ cases).
A slightly lower $\chi^2$ is obtained by optimizing the value of $t_0$.
However the unitarity bounds are still saturated in the optimized case with three (or higher) parameter fits, and it does not affect our result for $R(\eta_c)|_{\text{SM}}$.
We also checked that going beyond the BGL parameterization by imposing the additional constraints that $f_+$ and $f_0$ be maximal at zero recoil does not improve the bounds.
The only downside of the three parameter fits versus one or two parameter fits is the wider uncertainty band for $f_+$ as the recoil becomes larger.
The drawback is mild however as the differential partial width is kinematically suppressed at large recoil.
Furthermore, it could be ameliorated by an additional lattice measurement near maximum recoil.
This additional measurement could be synthesized using heavy-quark symmetry, see Section~\ref{sec:hqss}.

To obtain a prediction for $R(\eta_c)$ in the SM we insert our fitted values for the form factors into Eq.~\eqref{eq:gamma}, and integrate over $q^2$. We find
\begin{equation}
R(\eta_c)|_{\text{SM}} = 0.31^{+0.04}_{-0.02}. 
\end{equation}
The uncertainties are determined by simultaneously shifting the form factors in the numerator and denominator by $1\sigma$ from their respective central values.
Clearly, the uncertainty coming from the form factors mostly cancels in the ratio.
For the sake of the comparison, the relative uncertainty on the partial width involving a muon is approximately 30\%.
Our model-independent prediction is in good agreement with recent model calculations: $R(\eta_c)|_{\text{SM}} =$ $0.31^{+0.12}_{-0.07}$~\cite{Dutta:2017xmj}, $0.26$~\cite{Tran:2018kuv}, and $0.26 \pm 0.05$~\cite{Issadykov:2018myx}.
(For a calculation of the partial width for the decay $B_c \to \eta_c\, e\, \nu$ see~\cite{Ebert:2003cn}.)
Ref.~\cite{Berns:2018vpl}, which appeared on the arXiv shortly after our paper, also uses the BGL parameterization to obtain a model-independent prediction $R(\eta_c)|_{\text{SM}} = 0.29 \pm 0.05$.
Furthermore, our prediction agrees with those for $R(D)|_{\text{SM}}$~\cite{Lattice:2015rga, Na:2015kha, Bigi:2016mdz, Bernlochner:2017jka, Jaiswal:2017rve}, again suggesting that the dependence of the form factors on the flavor of the spectator quark is mild.

\section{Heavy-Quark Spin Symmetry and $B_c^+ \to J/\psi\, \ell^+\, \nu$}
\label{sec:hqss}
The results derived in Section~\ref{sec:ff} relied only on analyticity, crossing symmetry, and unitarity.
Form factors can generally be further constrained, or related to one another, if the system under consideration has additional symmetries.
Heavy-quark effective theory (HQET)~\cite{Georgi:1990um, Grinstein:1990mj} accurately describes hadrons with one heavy quark by treating the heavy quark as static, and expanding in powers of the heavy-quark mass, $m_Q$. 
The resulting heavy-quark symmetry allows one to describe all the form factors at low recoil for the decays $B^{(*)} \to D^{(*)} \ell \nu$ in terms of a single function, called the Isgur-Wise function, with fixed normalization~\cite{Isgur:1989vq, Isgur:1989ed}.
HQET is not well suited for hadrons containing more than one heavy quark.
The heavy quarks can exchange low energy gluons within the hadron leading to an IR-divergence that is regulated by the kinetic energy of the hadron.
The kinetic energy is formally suppressed in HQET by $1/m_Q$, thus a new effective field theory (EFT) with a new power counting is needed.
Non-relativistic QCD (NRQCD) is the appropriate EFT to use, and its expansion parameter is (not surprisingly) the speed of light.
Doubly-heavy hadrons in NRQCD possess a residual heavy-quark spin symmetry~\cite{Jenkins:1992nb, Colangelo:1999zn, Kiselev:1999sc} that, like heavy-quark symmetry, relates the form factors at zero recoil.
However HQSS does not fix the normalization of the form factors.

We use HQSS to relate the form factors for $B_c^+ \to \eta_c\, \ell^+\, \nu$ to $B_c^+ \to J/\psi\, \ell^+\, \nu$.
The differential partial width for $B_c^+ \to J/\psi\, \ell^+\, \nu$ is parameterized as follows
\begin{align}
\label{eq:gammapsi}
\frac{d\Gamma}{dq^2}(B_c^+ \to J/\psi\, \ell^+\, \nu) &= \frac{\eta_{ew}^2 G_F^2 |V_{cb}|^2 M_{B_c} \sqrt{\lambda_{\psi}}}{192 \pi^3} \left(1 - \frac{m_{\ell}^2}{q^2}\right)^2 \\
&\times \left[c_f f(q^2)^2 + c_1 \mathcal{F}_1(q^2)^2 + c_g g(q^2)^2 + c_2 \mathcal{F}_2(q^2)^2\right] \nn
\end{align}
with $\lambda_{\psi} = \lambda(q^2, M_{B_c}^2, M_{J/\psi}^2)$ and
\begin{equation}
c_f = \frac{2 q^2 + m_{\ell}^2}{M_{B_c}^4}, \quad c_1 = \frac{c_f}{2 q^2} , \quad c_g = \frac{\lambda_{\psi} c_f}{4} , \quad c_2 = \frac{3 \lambda_{\psi}}{8 M_{B_c}^4} \frac{m_{\ell}^2}{q^2} .
\end{equation}
The HQSS relations are~\cite{Kiselev:1999sc}
\begin{align}
\label{eq:hqss}
f_0(t_-) &= \frac{r}{1 + r} \frac{8}{3 + r - \sigma + r \sigma} f_+(t_-) , \\
f(t_{\psi-}) &= \frac{8 M_{B_c} r}{3 + r - \sigma + r \sigma} f_+(t_-) , \nn \\
g(t_{\psi-}) &= \frac{2 r}{M_{B_c} r_{\psi}} \frac{3 + \sigma}{3 + r - \sigma + r \sigma} f_+(t_-) , \nn \\
\mathcal{F}_2(t_{\psi-}) &= \frac{2 r}{r_{\psi}} \frac{3 + r_{\psi} - \sigma + r_{\psi} \sigma}{3 + r - \sigma + r \sigma} f_+(t_-) , \nn
\end{align}
where $t_{\psi-} = (M_{B_c} - M_{J/\psi})^2$,
\begin{equation}
r_{\psi} = \frac{M_{J/\psi}}{M_{B_c}}, \quad \sigma = \frac{m_c}{m_b} ,
\end{equation}
and as previously defined $r = M_{\eta_c} / M_{B_c}$, $t_- = (M_{B_c} - M_{\eta_c})^2$.
In addition, by construction we have~\cite{Boyd:1997kz}
\begin{equation}
\label{eq:F1f}
\mathcal{F}_1(t_{\psi-}) = M_{B_c} (1 - r_{\psi}) f(t_{\psi-}) .
\end{equation}
Our key assumption to make an estimate of $R(J/\psi)$ is that the last three relations in~\eqref{eq:hqss} and Eq.~\eqref{eq:F1f} hold beyond zero-recoil.
We then integrate Eq.~\eqref{eq:gammapsi} with respect to $q^2$.
A second estimate is obtained through the relation in the first line of~\eqref{eq:hqss}, whereby $R(J/\psi)$ becomes a function of $f_0$ via~\eqref{eq:hqss}, and these relations are again assumed to hold beyond zero recoil.
This gives us a range of predictions
\begin{equation}
R(J/\psi)|_{\text{SM}} = 0.26 \pm 0.02, 
\end{equation}
where the central value is the average of the two estimates, one using $f_0$ and one with $f_+$.
The uncertainty is taken to be the difference of the two estimates.
Our estimate is in good agreement with other recent determinations: $R(J/\psi)|_{\text{SM}} = $ $0.29 \pm 0.07$~\cite{Dutta:2017xmj}, $0.24 \pm 0.07$~\cite{Issadykov:2018myx} and, $0.283 \pm 0.048$~\cite{Watanabe:2017mip}. 
It is also in agreement with Ref.~\cite{Cohen:2018dgz}, which quotes the 95\% CL range 0.20 to 0.39.

\section{Conclusions}
\label{sec:con}
The hints of lepton flavor universality violation in the semileptonic decays of $B$ mesons should be explored as throughly as possible.
This includes studying the process $B_c^+ \to \eta_c\, \ell^+\, \nu$, which should be within the reach of LHCb at HL-LHC.
In this work we derived model-independent bounds on the form factors for $B_c^+ \to \eta_c\, \ell^+\, \nu$ using the BGL parameterization.
The bounds were obtained by fitting the form factors to the preliminary lattice data of the HPQCD Collaboration.
We then used the fitted form factors to compute the SM prediction for the ratio of branching fractions $R(\eta_c)$, yielding the result $R(\eta_c)|_{\text{SM}} = 0.31^{+0.04}_{-0.02}$.
Lastly, we related our fitted form factors for $B_c^+ \to \eta_c \, \ell^+\, \nu$ to those for $B_c^+ \to J/\psi\, \ell^+\, \nu$ using heavy-quark spin symmetry.
This allowed us to estimate the SM prediction for $R(J/\psi)$.
We found $R(J/\psi)|_{\text{SM}} = 0.26 \pm 0.02$ in good agreement with other determinations. 

\begin{acknowledgments}
We thank Ben Grinstein, Andrew Kobach, Henry Lamm, Richard Lebed, and Chien-Thang Tran for useful discussions.
This work was supported by the United States Department of Energy under Grant Contract DE-SC0012704.
\end{acknowledgments}

\bibliographystyle{utphys}
\bibliography{Bc_cc_ell_nu}

\end{document}